\documentclass[journal=jacsat,manuscript=article]{achemso}
\usepackage[version=3]{mhchem}

\usepackage{fullpage}
\usepackage[monochrome]{xcolor}
\usepackage{graphicx}
\usepackage[space]{grffile}
\usepackage{latexsym}
\usepackage{textcomp}
\usepackage{booktabs,array,multirow}
\usepackage{amsfonts,amsmath,amssymb,amsthm}
\usepackage{bm}
\usepackage{siunitx}
\usepackage{mathtools}
\usepackage{epstopdf}
\usepackage[normalem]{ulem}
\usepackage[sc,osf]{mathpazo}
\usepackage{tabularx}         
\usepackage{mciteplus}        
\usepackage{cprotect}

\usepackage[T1]{fontenc}
\usepackage[english]{babel}   

\AtBeginDocument{\singlespacing}

\PassOptionsToPackage{hyphens}{url}
\usepackage[colorlinks = true,linkcolor = blue,urlcolor = blue,citecolor = blue,anchorcolor = blue]{hyperref}
\usepackage[]{hyperref}
\usepackage{etoolbox}

\AtBeginDocument{\DeclareGraphicsExtensions{.pdf,.PDF,.eps,.EPS,.png,.PNG,.tif,.TIF,.jpg,.JPG,.jpeg,.JPEG}}

\author{Dhiman Ray\textsuperscript{$\Delta$}}
\affiliation{Atomistic Simulations, Italian Institute of Technology, Via Enrico Melen 83, 16152, Genova, Italy}
\author{Narjes Ansari\textsuperscript{$\Delta$}}
\affiliation{Atomistic Simulations, Italian Institute of Technology, Via Enrico Melen 83, 16152, Genova, Italy}
\author{Valerio Rizzi\textsuperscript{$\Delta$}}
\affiliation{Atomistic Simulations, Italian Institute of Technology, Via Enrico Melen 83, 16152, Genova, Italy}
\author{Michele Invernizzi}
\affiliation{Freie Universit\"at Berlin, Arnimallee 12, 14195, Berlin, Germany}
\author{Michele Parrinello}
\affiliation{Atomistic Simulations, Italian Institute of Technology, Via Enrico Melen 83, 16152, Genova, Italy}
\email{michele.parrinello@iit.it}

\title{Rare Event Kinetics from Adaptive Bias Enhanced Sampling}

\vspace{-1em}
	
\begin{document}

\begin{center}

\textsuperscript{$\Delta$}D.R., N.A. and V.R. contributed equally to the manuscript.
    
\end{center}
\begin{abstract}
\noindent
We introduce a novel enhanced sampling approach named OPES flooding for calculating the kinetics of rare events from atomistic molecular dynamics simulation. This method is derived from the On-the-fly-Probability-Enhanced-Sampling (OPES) approach [Invernizzi and Parrinello, JPC Lett. 2020], which has been recently developed for calculating converged free energy surfaces for complex systems. In this paper, we describe the theoretical details of the OPES flooding technique and demonstrate the application on three systems of increasing complexity: barrier crossing in a two-dimensional double well potential, conformational transition in the alanine dipeptide in gas phase, and the folding and unfolding of the chignolin polypeptide in aqueous environment. From extensive tests, we show that the calculation of accurate kinetics not only requires the transition state to be bias-free, but the amount of bias deposited should also not exceed the effective barrier height measured along the chosen collective variables. In this vein, the possibility of computing rates from biasing suboptimal order parameters has also been explored. Furthermore, we describe the choice of optimum parameter combinations for obtaining accurate results from limited computational effort. 
\end{abstract}

\newpage
\section*{Introduction}

The use of atomically detailed molecular dynamics simulations has become pervasive in most scientific fields. For this reason, much effort has been devoted to extend their capability. One outstanding limitation is the restricted time scale that can be reached. Often, the reason for wanting to extend the time scale exploration can be linked to the presence of different metastable states separated by large free energy barriers. Examples include chemical reactions or protein conformational changes. Such transitions can be characterized as rare events since large free energy barriers make the timescale of state-to-state transitions much larger than the one that is accessible to conventional molecular dynamics simulations. This makes sampling difficult. To remedy this difficulty, a large variety of enhanced sampling methods have been proposed~\cite{Ahmad2022}. In these methods the natural dynamics of the system is altered to accelerate sampling. Most often this is done by adding an external bias. Enhanced sampling methods allow computing static properties such as the free energy but, since they distorts dynamics, they make it challenging to recover the correct kinetics \cite{Palacio-Rodriguez2021,donati2018girsanov}. In an alternative approach, path sampling methods focus on the generation of reactive paths from which rates can be computed \cite{Bolhuis2002,huber1996weighted,zhang2010weighted,zuckerman2017weighted}. Another class of path sampling techniques discretize transition paths using high dimensional interfaces in an attempt to obtain converged kinetics from short independent simulations limited in different regions of the configurational space~\cite{Faradjian2004,West2007,allen2006simulating,Ray2020WeightedSimulations,Ray2022}. The advantage of path sampling is that no external bias needs to be applied, which ensures minimal perturbation to the natural dynamics of the system. But the determination of static properties can be more challenging due to reduced exploration.  
 
Some time ago, Grübmüller and Voter~\cite{Grubmuller1995,Voter1997} noted that in biased runs, dynamical information can still be recovered if one takes care of not adding bias to the transition state region. The first approach is called conformational flooding \cite{Grubmuller1995}, while the second goes under the name of hyperdynamics \cite{Voter1997}. In the last decade, our group has made several attempts at exploiting Grübmüller and Voter's suggestion in the context of the collective variable based enhanced sampling methods that were being developed. The first effort was based on Metadynamics~\cite{Laio2002,Barducci2008}, a method that, like the ones that followed, is based on an on-the-fly determination of a bias that is a function of a few collective variables (CVs) and is periodically updated. This first such method goes under the name of Infrequent Metadynamics~\cite{Tiwary2013} and relies on the fact that the system spends very little time transiting between metastable states, and consequently it is unlikely that the bias is applied in saddle points if the frequency of bias deposition is very low. This approach has found several successful applications~\cite{Casasnovas2017,Tiwary2015Trypsin,ribeiro2018reweighted,rizzi2019onset,mondal2018atomic,Kulkarni2021BPTI,yang2022using}. However, due to the requirement that the bias deposition rate is low, the convergence of this approach is slow, and the criteria used to assess the validity of the result can some time fail~\cite{Dickson2019}. It also requires, which is not a surprise, the use of good CVs. More recently, we have introduced along similar lines, two other techniques: the Variationally Optimized Free-Energy Flooding (VES-flooding)~\cite{McCarty2015} and Gaussian Mixture Based Enhanced Sampling (GAMBES)~\cite{Debnath2020}, both of which have been successfully applied to both chemical and biological systems \cite{Piccini2017,palazzesi2017conformational,debnath2022computing}. Despite these developments, recovering the kinetics in practical problems remains a significant challenge due to the complexity, and the high dimensionality of such systems.

In this paper we want to ameliorate these techniques, by using the recently developed On-the-fly Probability Enhanced Sampling (OPES)~\cite{Invernizzi2020} which has shown to be a significant improvement over Metadynamics, and it is easier to use than the variational approach. Besides its efficiency, OPES shares with the variational approach the possibility of setting an upper bound to the value of the bias that is deposited. This has led to a modified OPES protocol that can be used to calculate rare events rates and that we refer to as OPES flooding (OPES$_\mathrm{f}$). In the subsequent sections we describe the theoretical and computational details of this method and show its application to a 2D toy model system, the conformational transition in alanine dipeptide, and the folding and unfolding of the chignolin mini-protein. We discuss carefully the advantages and limitations of this approach and direct the reader on how to choose the optimal set of parameters to extract good kinetic properties with a limited amount of computational effort.  


\section*{Methods}
\subsection*{Theory}
The OPES flooding approach stems from its predecessors OPES \cite{Invernizzi2020} and the Variationally Optimized Free-Energy Flooding \cite{McCarty2015} techniques. In these adaptive biasing methods, an external bias is added to the potential energy $U(\mathbf{R})$ of the system. The external bias $V(\mathbf{s})$ is applied along a set of collective variables (CVs) ($\mathbf{s}(\mathbf{R})$) which is a function of the atomic coordinates $\mathbf{R}$ and serves as a low dimensional representation of the slowest degrees of freedom involved in the rare event process. The free energy of the system, when expressed as a function of the CVs, is called the Free Energy Surface (FES) and is given by $F(\mathbf{s}) = -(1/\beta)\ln P(\mathbf{s})$ where $\beta = 1/k_BT$, the inverse Boltzmann factor and $P(\mathbf{s})$ is the marginal probability distribution in the CV space, which is given by $P(\mathbf{s}) \propto \int \mathrm{d}\mathbf{R} \exp(-\beta U(\mathbf{R})) \delta [\mathbf{s} -\mathbf{s}(\mathbf{R})]$. The bias $V(\mathbf{s})$ is periodically updated during the simulation, facilitating the exploration of the configurational landscape of the system or the calculation of the free energy surface. The pioneering approach in this category is  Metadynamics (MetaD) \cite{Laio2002,Barducci2008} where Gaussian hills are deposited along the CV to iteratively build the biasing potential $V(\mathbf{s})$. 


OPES is an evolution of MetaD where Gaussian kernels are used for reconstructing the marginal probability distribution along the CVs $P(\mathbf{s})$ rather than for directly building the bias potential.
This allows to directly take advantage of tools from the kernel density estimation literature and leads to fewer input parameters and faster convergence. OPES can be used to sample a variety of target distributions~\cite{Invernizzi2020d,Invernizzi2022}, i.e. the marginal distribution in the CV space in presence of the external bias: $p^{\text{tg}}(\mathbf{s}) \propto \int \mathrm{d}\mathbf{R} \exp(-\beta (U(\mathbf{R}) + V(\mathbf{s})) \delta [\mathbf{s} -\mathbf{s}(\mathbf{R})]$.
A typical choice akin to MetaD is the Well-Tempered distribution $p^{\text{tg}}(\mathbf{s}) \propto [P(\mathbf{s})]^{1/\gamma}$, where the bias factor $\gamma > 1$ is a parameter that controls the target distribution smoothness compared to the unbiased one.
The bias is then obtained in a self-consistent manner from the on-the-fly probability estimates, so that, at convergence, it is:
\begin{equation}
    V(\mathbf{s})=-\frac{1}{\beta} \log \frac{p^{\text{tg}}(\mathbf{s})}{P(\mathbf{s})}\, .
\end{equation}
To ensure stability and speedup convergence, two small correction are added to this simple formula.
The first one is a regularization term $\epsilon=e^{-\beta \Delta E/(1-1/\gamma)}$, that ensures that the bias is always bound to a maximum absolute value of $\Delta E$, referred to as the \textit{barrier} parameter (see SI of Ref.~\citenum{Invernizzi2020}).
The second one, $Z$, is a normalization over the explored volume of the CV space that is estimated on-the-fly in a way detailed in Refs.~\citenum{Invernizzi2020,Invernizzi2022}.
With these modifications, the iterative OPES bias for well-tempered target at step $n$ is written as:
\begin{equation}
    V_n(\mathbf{s})=(1-1/\gamma)\frac{1}{\beta} \log \left( \frac{P_n(\mathbf{s})}{Z_n}+\epsilon \right)\, ,
\end{equation}
where $P_n(\mathbf{s})$ is the weighted kernel density estimation of the unbiased $P(\mathbf{s})$ at step $n$.
The kernel bandwidth used for this estimate shrinks as the simulation proceeds, allowing for a finer and finer approximation.
This is possible also thanks to a kernel merging algorithm that serves the twofold goal of keeping the bias evaluation efficient and reducing the bandwidth of already deposited kernels.
We again refer to Ref.~\citenum{Invernizzi2020} for all the details about how the probability estimate is obtained.

The OPES technique can efficiently evaluate the biasing potential, which we propose to utilize in the realm of free energy flooding approach to recover the kinetics of the underlying dynamics. The free energy flooding method is derived from the well known conformational flooding approach~\cite{Grubmuller1995} for computing kinetics from biased molecular dynamics simulation. In this technique, the free energy basin is filled up to a predefined level by depositing bias using one of many enhanced sampling approaches including variationally enhanced sampling (VES) \cite{McCarty2015}, Infrequent Metadynamics \cite{Tiwary2013} or accelerated molecular dynamics (aMD) \cite{Hamelberg2004}. This helps in increasing the sampling in the initial state basin as well as reducing the effective activation barrier for transiting into new conformational states. The kinetics of the process is recovered from the flooding simulations using the key assumption that no bias is deposited in the transition state and the properties of the transition state remain unaltered even after applying the flooding in the free energy basin(s). Under this condition, the ratio between the true mean first passage time (MFPT) ($t$) and the MFPT in the flooding simulation ($t_\mathrm{f}$) is given by 
\begin{equation}
    \label{eqn:time-rescaling-1}
    \frac{t}{t_f}= \left\langle \exp(\beta V_\mathrm{f}(\mathbf{R})) \right\rangle_{U+V_\mathrm{f}}
\end{equation}

where $V_\mathrm{f}(\mathbf{R})$ is the bias applied in the flooding simulation as a function of the atomic coordinates $\mathbf{R}$. The ensemble average is computed from the flooding trajectory in the presence of both the potential energy of the system $U$ and the bias potential $V_\mathrm{f}$. This ratio is referred to as the acceleration factor, which is a measure of computational efficiency of the flooding based approach in estimating the rate constant. It is important to note that the right hand side of Eq.~\ref{eqn:time-rescaling-1} is an ensemble average in presence of the flooding bias $V_\mathrm{f}$ evaluated within the initial state minimum. This necessitates that the bias is converged (i.e. it reaches a quasi static value) in the initial state minimum. The importance of this condition  is illustrated in our calculations. Ideally, the MFPT $t$ follow a Poissonian distribution $\frac{1}{\tau}\mathrm{e}^{-t/\tau}$ where $\tau$ is the characteristic time associated to the studied rare event. By fitting a Poisson distribution to the observed $t$, one can estimate $\tau$ and a common tool to assess the quality of this estimation is the Kolmogorov-Smirnov (KS) test with its p-value~\cite{Salvalaglio2014}.

In the OPES flooding approach, the flooding bias $V_\mathrm{f}$ is built using OPES which allows the user to have finer control over the level up to which the initial state basin is filled with bias in comparison to infrequent metadynamics. Alongside, we introduce an excluded region $\chi_{\textrm{exc}}(s)$ where kernels are not deposited, so that the bias there will remain uniform. For example, the excluded region can be defined as $\chi_{\textrm{exc}}(s) = \mathbb{H}(s-s_\mathrm{exc})$ where $\mathbb{H}$ is Heaviside step function, $s$ is the CV and $s_\mathrm{exc}$ is the boundary of the excluded region in the CV space. In such a scenario, no Gaussian kernel will be deposited by OPES if the center of the kernel is located at a CV value above $s_\mathrm{exc}$. See SI for discussion on the dimensionality of $s_\mathrm{exc}$. Using a combination of the barrier parameter $\Delta E$ and the excluded region $\chi_{\text{exc}}(s)$, we can precisely avoid biasing the transition state while filling up the initial state basin to a significant extent to accelerate the transition from the initial state to the final state. 


\begin{figure}[H]
    \centering
    \includegraphics[width=0.7\textwidth]{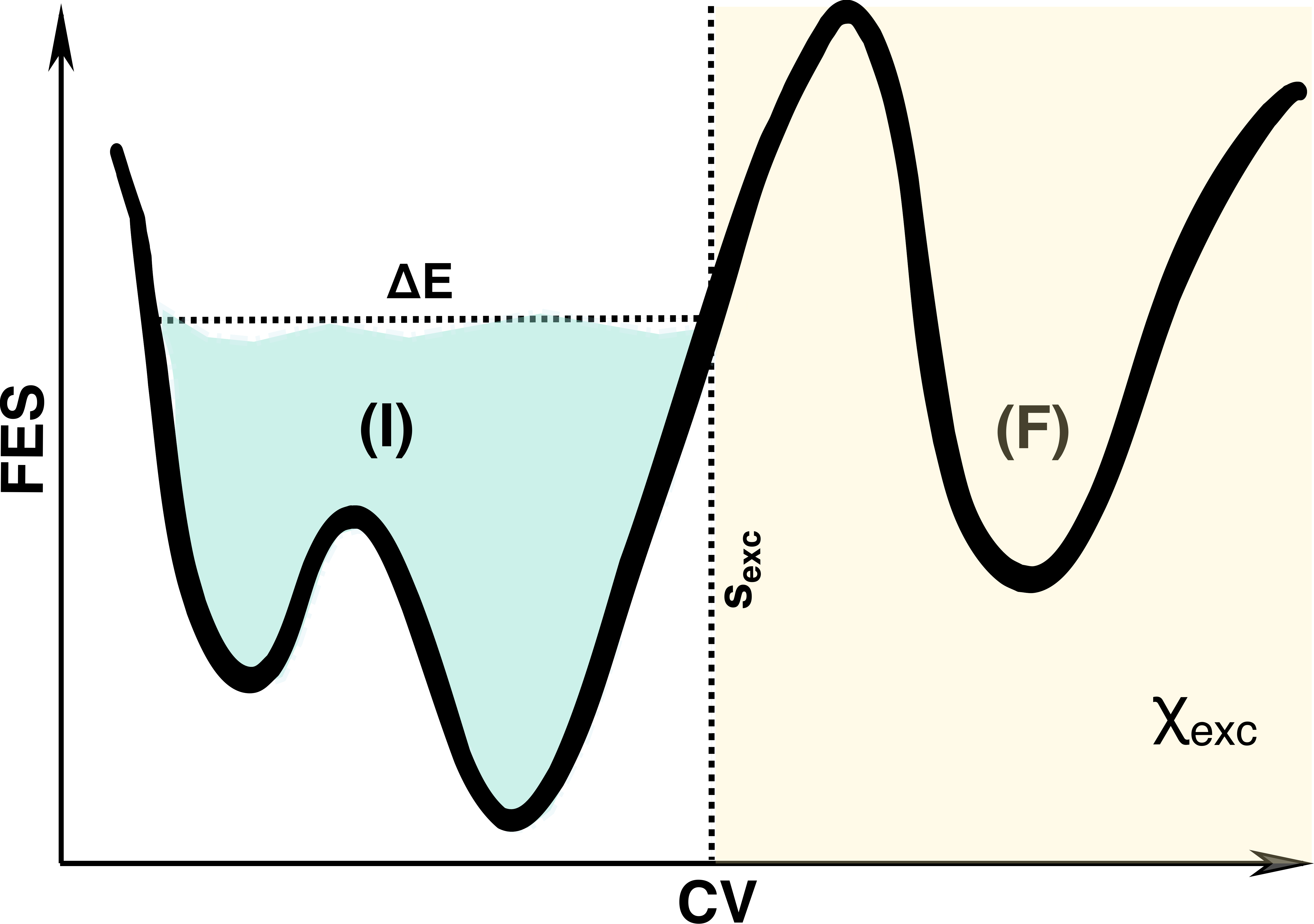} 
    \cprotect\caption{A schematic representation of the OPES flooding approach. The initial state (I) basin is filled with external bias (colored light green) up to a predefined level controlled by the barrier parameter $\Delta E$. The excluded region $\chi_{\textrm{exc}}(s)$ denotes the area in which no bias is deposited to avoid the transition state. }
    \label{fig:schematic}
\end{figure}

\subsection*{Computational Details}

The OPES flooding method is available in PLUMED from version 2.8, through the contributed OPES module. We studied the applicability of this approach for calculating the kinetics of three different rare event processes: barrier crossing in 2D toy model of modified Wolfe-Quapp potential, the conformational transition in gas phase alanine dipeptide, and the folding and unfolding of chignolin miniprotein in explicit solvent.

\textbf{Modified Wolfe-Quapp Potential:} The OPES flooding approach is first tested on the two-dimensional toy model system of the modified Wolfe-Quapp potential\cite{Invernizzi2019}. The functional form of the potential energy surface is the following:

\begin{equation}
    U(x,y) = 2(x^{4}+y^{4}-2x^{ 2}-4y^{ 2}+2xy+0.8x+0.1y+9.28)
\end{equation}

We are interested in investigating different levels of suboptimality of the CV. For convenience, instead of changing the CV, we will always use the $x$ coordinate but rotate the plane  counterclockwise around the origin by an angle of $\theta$, as shown in Fig. 3. This transforms the coordinate frame in the following manner: 
\begin{equation}
\begin{split}
     x' = \cos(\theta)x - \sin(\theta)y \\
     y' = \sin(\theta)x + \cos(\theta)y
\end{split}
\end{equation}

For simplicity, we always denote the $x'$ coordinate as $x$.
We chose $\theta = 27^{\circ}, \ 54^{\circ}$ and $90^{\circ}$ with increasing quality of the CV in distinguishing the two minima. OPES flooding simulations were performed with barrier parameters ranging from 4 $k_\mathrm{B}T$ to 16 $k_\mathrm{B}T$ (in 2 $k_\mathrm{B}T$ intervals) to cover regions sufficiently above and below the free energy barrier in the three systems. The unit of the energy is chosen to be $k_\mathrm{B}T$. The boundary of the excluded region is varied between -1.0 to 0.0 at an interval of 0.25 for all the $\Delta E$ values, to determine the optimum choice of these two settings in terms of obtaining an appropriate kinetics. For each combination, 50 Langevin dynamics simulations were performed  with a damping coefficient of 10/timeunit with 0.002 timeunit timestep using the \verb|ves_md_linearexpansion| action of the PLUMED software. The trajectories were stopped when they reached $x > 1.6$ using the COMMITTOR command of PLUMED. 

\begin{figure}[H]
    \centering
    \includegraphics[width=0.7\textwidth]{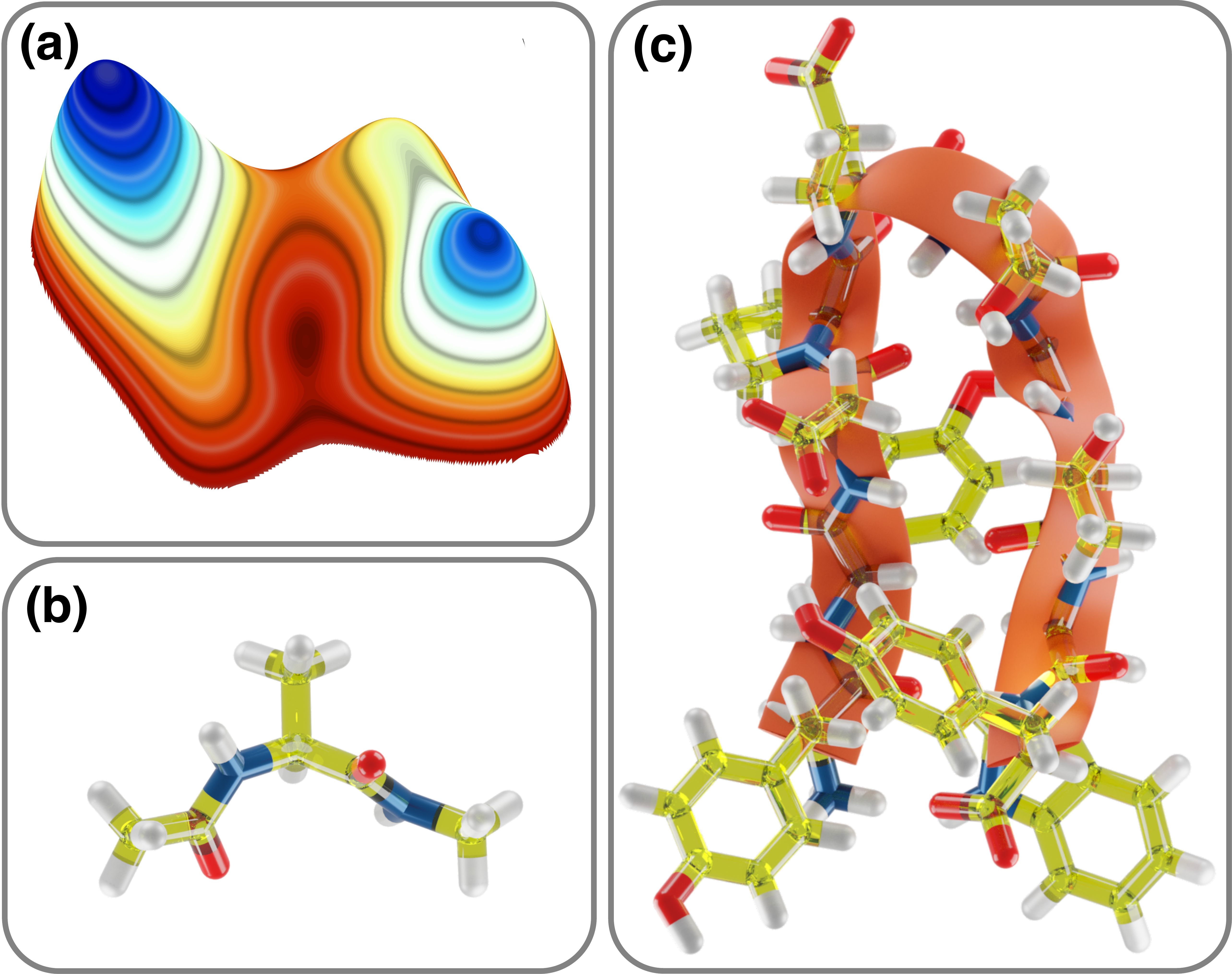} 
    \cprotect\caption{A cartoon representation of (a) modified Wolfe-Quapp potential, (b) alanine dipeptide and (c) chignolin. }
    \label{fig:sys}
\end{figure}

\textbf{Alanine Dipeptide:}
Next we tested the OPES flooding scheme on the system of alanine dipeptide in the gas phase which is the smallest molecular system that captures all features of a rare event and has been used widely for testing various enhanced sampling methods. The 22-atom system has been modeled using the AMBER99SB-ILDN force field. The molecular dynamics simulations were performed using the GROMACS 2021.5 \cite{Abraham2015} package patched with the development version of PLUMED 2.9 \cite{Tribello2014}. All simulations were propagated starting from the $C7eq$ state of alanine dipeptide in order to observe transitions from $C7eq$ state to $C7ax$ state \cite{Valsson2016}. A long unbiased simulation was also propagated for $\sim$ 80 $\mu$s to compare the accuracy of the predicted results from the OPES flooding approach. 

In case of the OPES flooding simulations, the bias has been applied only to the $\phi$ torsion angle. The barrier parameter $\Delta E$ has been varied between 15 and 40 kJ/mol with an interval of 5 kJ/mol, and the boundary of the excluded region is varied from -0.7 radians to -0.1 radians with an interval of 0.1 radians along the $\phi$ collective variable. Trajectories were stopped when they reached the $C7ax$ state, i.e. when $\phi > 1.0$ rad. 

To check the effect of the choice of CV on the kinetics, another set of OPES$_\mathrm{f}$ simulations were performed with the bias applied along the $\psi$ torsion angle. Three different $\Delta E$ values (5, 7.5 and 10 kJ/mol) have been tested in this case with only one exclusion boundary at $\psi = 0.0$ radian. The choice of these parameters are described later in the manuscript.

\textbf{Chignolin miniprotein:}
The final system that we studied in this work is the CLN025 mutant of Chignolin miniprotein \cite{Honda2008} in explicit solvent environment. The folding and unfolding of this 10-residue small protein has been previously investigated using long unbiased MD simulation~\cite{Lindorff-Larsen2011}. We perform our simulations with identical conditions to compare our results with the brute force MD dataset. The protein has been solvated in a box of 1907 water molecules; 2 Na+ ions have been added to neutralize the system. We modeled the protein using the CHARMM22$^*$ force field \cite{piana2011robust} and the solvent has been modeled by the CHARMM TIP3P force field \cite{mackerell1998all}. 

We used three different CVs for the OPES$_\mathrm{f}$ simulations: the Harmonic Linear Discriminant Analysis (HLDA) CV based on 6 interatomic contacts within the protein \cite{Mendels2018a}, a Deep Linear Discriminant Analysis (Deep-LDA) CV \cite{Bonati2020} trained on the 210 descriptor sets proposed by Bonati et al. \cite{Bonati2021} and a (Deep Time-lagged Independent Component Analysis (Deep-TICA) CV \cite{Bonati2021} trained on the same set of descriptors from the long unbiased trajectory~\cite{Lindorff-Larsen2011}. The details of the training of Deep-LDA and Deep-TICA CVs have been described in the Supporting Information. The choice of $\Delta E$ and $\chi_{\textrm{exc}}(s)$ for each case is discussed in detail later in the manuscript. For each set, 20 independent OPES$_\mathrm{f}$ simulations were propagated to study folding and unfolding. The trajectories were initiated from the same folded  or unfolded structures, but with different initial velocities; they were stopped when an unfolding or folding event is observed.

\section*{Results}
\subsection{Modified Wolfe-Quapp potential}
\begin{figure}[H]
    \centering
    \includegraphics[width=0.85\textwidth]{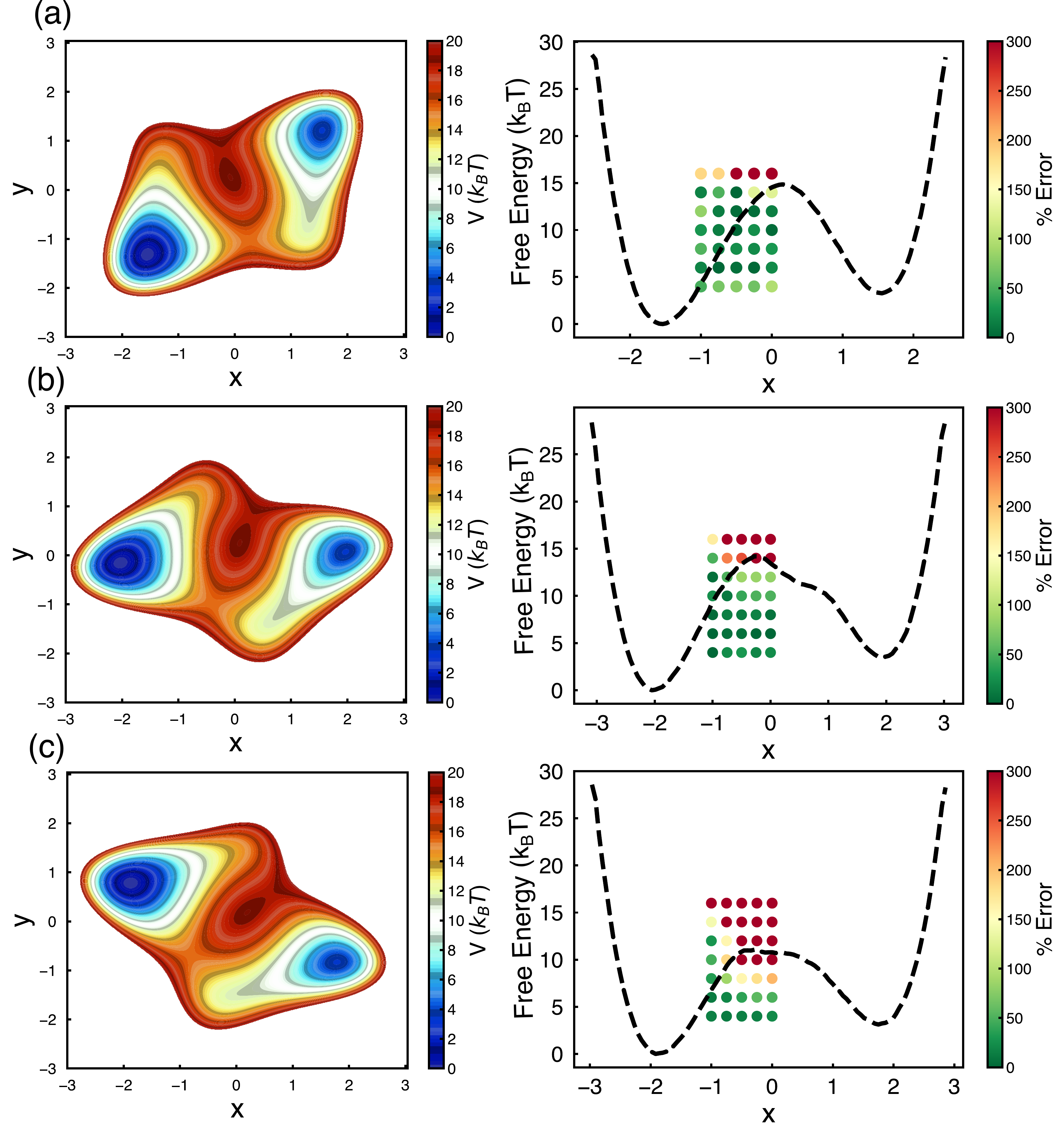}. 
    \cprotect\caption{The results of the application of the OPES$_\mathrm{f}$ technique on the modified Wolfe-Quapp potential. The potential energy surface for the three systems tested are shown in the left panel (a) $\theta = 90^{\circ}$, (b) $\theta = 54^{\circ}$, and (c) $\theta = 27^{\circ}$. In the right panel we plot the FES along $x$. In the same panel we measure the quality of the results obtained using different combinations of $s_\textrm{exc}$ and $\Delta E$, where $s_\textrm{exc} \in \{-1.0, -0.75, -0.5, -0.25, 0.0\}$ and $\Delta E \in \{4,6,8,10,12,14\}$ in $k_BT$ unit. The coordinate of each filled circle represents a specific parameter combination ($s_\textrm{exc}$, $\Delta E$) tested in this work. The color of the circles corresponds to the mean absolute error of the $\tau$ computed from OPES$_\mathrm{f}$ with reference to the estimate from unbiased trajectories. Each filled circle represents results obtained from 50 independent simulations. }
    \label{fig:2d_model_results}
\end{figure}

The simplicity of the modified Wolfe-Quapp potential allowed us to perform multiple tests with various combinations of $\Delta E$, $\chi_{\textrm{exc}}(s)$, and collective variables. The results are summarized in Fig. \ref{fig:2d_model_results}. One of the key observations is that the correct transition timescale can be recovered from the OPES$_\mathrm{f}$ simulation with any of the three CVs used in this work. However, the use of good quality CVs allowed a wider choice of hyperparameters to be used without compromising the quality of the results. The best quality CV (i.e. $\theta=90^{\circ}$ that makes the CV distinguish best the basins and the transition state region) predicts the transition timescales from the left basin to the right basin with reasonable accuracy, for any combination of $\Delta E$ between $4-14 \ k_\mathrm{B}T$ and $s_\textrm{exc}$ between $-1.0$ and $0.0$ . The results become poorer only when we choose a $\Delta E$ higher than the actual height of the free energy barrier for the transition. It is noteworthy that exceeding the free energy barrier even by a slight amount can significantly decrease the accuracy of the kinetic estimates, and one should be very careful in choosing the barrier parameter. Of course, this should have been expected since it corresponds to overfilling the basin.

On the other hand, in case of poorer CVs such as $\theta = 27^{\circ}$ and $54^{\circ}$ where the x-axis does not align well with the transition path, the correct kinetics can still be recovered but only using a lower value of $\Delta E$. The choice of the barrier has to be lower than the actual free energy barrier along the chosen CV. As a result we can get acceptable kinetics ($<200 \ \%$ absolute error) when $\Delta E$ is less than or equal to $8 \ k_\mathrm{B}T$ and $12 \ k_\mathrm{B}T$ for the $\theta = 27^{\circ}$ and $54^{\circ}$ case respectively. Irrespective of the quality of the CV, the $\chi_{\textrm{exc}}(s)$ needs to be chosen carefully to avoid depositing bias in the transition state, except for very low $\Delta E$ values where the extent of exploration of the conformational space is determined exclusively by the level of filling of the free energy minima. One might think that choosing a very low barrier is the best option to obtain reliable results. However, this choice would significantly reduce the calculation efficiency as the acceleration factor may reduce by orders of magnitude when reducing $\Delta E$. Also, the $s_\textrm{exc}$ should be kept far from the transition state to avoid the possibility of the tail of the deposited Gaussian kernels corrupting the transition state.

It is important to note, as discussed above, that avoiding bias deposition on the transition region is necessary but not sufficient to calculate appropriate kinetic properties. Irrespective of the location of the excluded region, if the amount of bias deposited in the initial state basin is higher than the activation energy barrier the system never reaches the quasistatic regime before the transition event. As a result, the ensemble average of $\exp(-\beta V_f)$ (see Eq. \ref{eqn:time-rescaling-1}) fails to converge, leading to poorer estimate of the kinetics. This phenomenon is demonstrated explicitly in the next example of alanine dipeptide. 

\subsection{Alanine dipeptide}
\begin{figure}[H]
    \centering
    \includegraphics[width=1\textwidth]{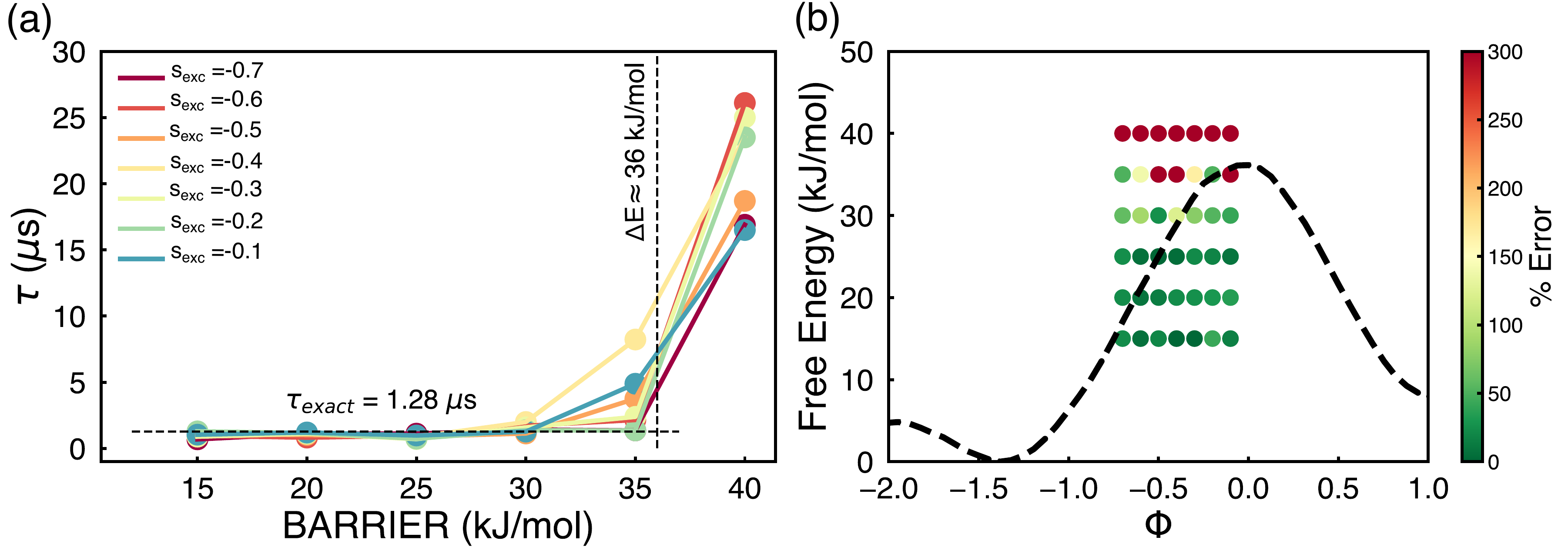}. 
    \cprotect\caption{Results of OPES$_\mathrm{f}$ simulation of the alanine dipeptide. (a) The variation of the transition timescale $\tau$ with $\Delta E$ for different excluded regions. The value of the activation energy (36 kJ/mol) has been indicated by a vertical dashed line. (b) Plot of the accuracy of the OPES$_\mathrm{f}$ kinetics projected on to the FES along $\phi$ dihedral angle. The color scheme is identical to Fig. \ref{fig:2d_model_results}.} 
    \label{fig:ala2_phi_results}
\end{figure}

\begin{figure}[H]
    \centering
    \includegraphics[width=0.9\textwidth]{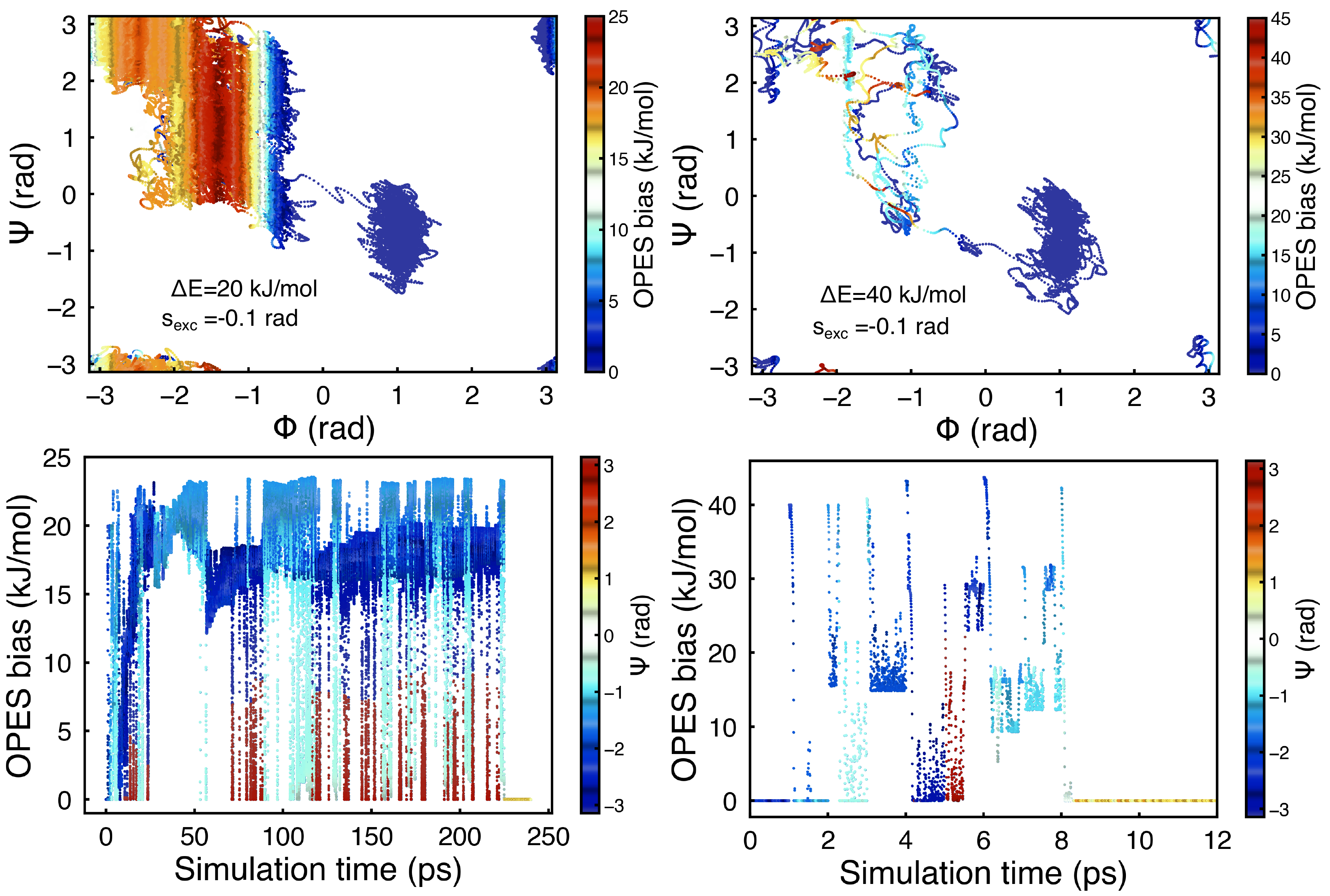}. 
    \cprotect\caption{Upper panel: When $\Delta E$ is substantially lower than the activation barrier (left), the OPES bias converges in the initial state minimum before transiting into the final state. When $\Delta E$ is higher than the activation barrier (right), the OPES bias fails to converge in the initial state minimum producing poor estimate of the kinetics. Bottom panel: time evolution of the OPES bias as a function of simulation time, plotted until the transition event. Each figure in the bottom panel correspond to the same trajectories in the immediate upper subplot. } 
    \label{fig:ala2_bias_convergence}
\end{figure}

From the OPES$_\mathrm{f}$ simulations of gas phase alanine dipetide, with various different parameter combinations, we uncovered very similar trends. When the $\Delta E$ is less than the activation barrier in the free energy surface, the predicted kinetics is in excellent agreement with the result obtained from long unbiased simulation ($\tau$ = 1.28 $\mu$s from 57 transitions) (Fig. \ref{fig:ala2_phi_results}). Unlike the 2D  model, the absolute error in the kinetics is very sensitive to $s_\mathrm{exc}$ and the p-values of the 2 tailed KS test are not well correlated with the accuracy of the predicted rates, making p-values a relatively less reliable approach to quantify the quality of the rate estimation. The key observation is that the bias needs to be converged inside the initial state basin before the transition to final state, in order to achieve a better accuracy. A representative trajectory for each situation (converged bias vs unconverged bias) is depicted in Fig. \ref{fig:ala2_bias_convergence}. When the bias is not converged before the transition, the system is driven out of equilibrium via the application of external force making the dynamics unphysical and consequently unsuitable for estimating kinetic properties. 

\textcolor{red}{It is interesting to note that  within the $C7_{eq}$ state there is a small energy barrier along the $\psi$ torsion angle, which has previously been indicated as the cause of the erroneous estimation of transition rates by infrequent metadynamics \cite{Dickson2019}. In the OPES flooding approach, even when just the $\phi$ angle is used as the CV, the estimated rates are unaffected by this extra hidden barrier, as the accuracy of the result depends only on the convergence of bias in the initial state minimum and consequent accurate evaluation of the ensemble average of $\exp(\beta V_\mathrm{f}(\mathbf{R}))$. However, the applicability of OPES$_f$ in systems with multiple high free energy barriers or in rugged free energy landscapes remains to be explored, and we intend to investigate these in future work.  }

To understand the potential role of the choice of collective variable on the accuracy of the predicted kinetics, additional OPES$_\mathrm{f}$ simulations have been performed with bias applied along the $\psi$ degree of freedom. In the alanine dipeptide $\psi$ stands as an example of a poor CV to observe the transition between $C7_{eq}$ and $C7_{ax}$ configurations as it can hardly distinguish between the two states let alone the transition state. The activation barrier in the free energy surface along $\psi$ ($\Delta G^{\dagger}_{\text{estimate}}$) is $\sim$ 9 kJ/mol as opposed to true activation barrier ($\Delta G^{\dagger}_{\text{true}}$) of $\sim$ 36 kJ/mol along $\phi$ torsion angle. As a result we were forced to choose $\Delta E$ to be less than 10 kJ/mol significantly reducing the efficiency of our calculation. The setup of our test is summarized in supporting information (Fig. S1). In all three biasing schemes, the predicted kinetics is in reasonable agreement with the timescales obtained from unbiased simulations. Although the accuracy was slightly poorer compared to the case of $\phi$ CV, they are well within the acceptable range for this type of calculation. This is a significant outcome, reinforcing the fact that the result of the predicted kinetics from OPES$_\mathrm{f}$ simulations are not very sensitive to the choice of CV.  

Such an observation have previously been discussed by Tiwary and Parrinello when they introduced the first version of the infrequent metadynamics method \cite{Tiwary2013} but we are able to explicitly demonstrate it in the current work. This indicates that, although the acceleration factor (i.e. efficiency) of the calculation can significantly improve with increasing the quality of CV, it is possible to achieve a correct estimate of the kinetics even when identification of a very good CV is extremely challenging. \textcolor{red}{This surprising result stems from the fact that even if the CV is not good and the exclusion zone extends into the TS region, the transition state can still be bias-free if the $\Delta E$ is sufficiently small. If we choose our $\Delta E$ to be lower than $\Delta G^{\dagger}_{\text{estimate}}$, the TS region will remain bias-free irrespective of the location of $s_{\text{exc}}$. However, in this situation the acceleration factor is poor, which is a trade-off for using a poor collective variable.}
\begin{table}[]
\caption{Results of the OPES$_\mathrm{f}$ simulation for alanine dipeptide, applying bias on the $\psi$ torsion angle.}
\begin{tabular}{|l|l|l|l|l|l|}

\hline
\textbf{$\Delta E$ (kJ/mol)} & \textbf{$\tau$ ($\mu$s)} & \textbf{p-value}    & $\mu$ ($\mu$s) &$\sigma$ ($\mu$s)  & \textbf{Acceleration factor} \\ \hline
\textbf{Unbiased}  & 1.28       & 0.987    & 1.17 & 0.98             & -          \\ \hline
5.0      & 1.56      & 0.97 & 1.51 & 1.38 & 7.2        \\ \hline
7.5  & 3.29      & 0.94 & 3.75 & 4.16 & 31.3        \\ \hline
10.0 & 2.03      & 0.69 & 1.89 & 1.78 & 36.3        \\ \hline
\end{tabular}
\end{table}

\subsection{Chignolin miniprotein}
After the successful application of OPES flooding approach on model systems, we studied the kinetics of the more practical example of protein folding/unfolding in explicit solvent.
The choice of $\Delta E$ and $\chi_{\textrm{exc}}(s)$ were made from the FES obtained from the 106 $\mu$s long unbiased trajectory. This helped us to gauge the optimum parameters for our simulations to improve the acceleration factor without sacrificing the accuracy of the predicted kinetics. But such an approach is not practical for generic systems for which a converged free energy surface may not be available beforehand. In such situations an approximate idea of the optimum combination of $\Delta E$ and $s_\textrm{exc}$ can be obtained from converging the free energy profile within the initial state basin confining the exploration of configurational space using a half-harmonic wall. This is possible thanks to the fact that these parameters are only dependent on the FES of the initial state and the barrier height. The process of obtaining approximate $\Delta E$ and $\chi_{\textrm{exc}}(s)$ is described with example in the supporting information.   

The timescales of the unfolding process obtained from the OPES$_\mathrm{f}$ method with three different ML based CVs are in reasonable agreement with the results from the unbiased simulation by Lindorff-Larsen et al. \cite{Lindorff-Larsen2011}. The agreement is particularly good for the Deep-LDA and Deep-TICA CV, although these CVs have been trained on numerous intramolecular features in a relatively black box manner. The acceleration factor obtained from the deep NN based CVs are higher than the HLDA CV, which reduces the overall computational cost in estimating the kinetic properties. Despite being developed for calculating free energy surfaces, our results indicate that the deep learning collective variables are superior to the manually selected collective variables also in kinetics calculations. 

Increasing $\Delta E$ by a few kJ/mol to reach near the activation barrier alters the predicted transition times by a factor of 2-3 while increasing the acceleration factor significantly. Although the accuracy of kinetics can be improved with a very conservative bias deposition, for computationally intensive problems, an approximate rate constant can still be calculated using a $\Delta E$ closer to the activation barrier (See SI Fig. S2).

\begin{table}[]
\caption{Unfolding kinetics of chignolin from OPES$_\mathrm{f}$ simulation. }
\label{tab:Unolding}
\begin{tabular}{|l|l|l|l|l|l|l|}

\hline
\textbf{Method/CV} & \textbf{$\tau$ ($\mu$s)} & \textbf{p-value}    & $\mu$ ($\mu$s) &$\sigma$ ($\mu$s)  & \textbf{Acceleration factor} & $\Delta E$ \textbf{(kJ/mol)}\\ \hline
\textbf{Unbiased}  & 2.20       & -    & -      & -       & -  & -        \\ \hline
\textbf{HLDA}      & 6.33      & 0.96 & 6.02 &5.58 & 115    & 15    \\ \hline
\textbf{Deep LDA}  & 2.93      & 0.71 & 3.18 &3.79 & 149    & 15    \\ \hline
\textbf{Deep TICA} & 3.21      & 0.56 & 4.87 &8.22 & 183    & 15    \\ \hline
\end{tabular}
\end{table}

\begin{table*}[h!tb]
\centering
\caption{Folding kinetics of chignolin from OPES$_\mathrm{f}$ simulation.}
\label{tab:Folding}
\begin{tabular}{|l|l|l|l|l|l|l|}
\hline
\textbf{Method/CV} & $\tau$ ($\mu$s)  & \textbf{p-value } &  $\mu$ ($\mu$s) &$\sigma$ ($\mu$s) &  \textbf{Acceleration factor} & \textbf{$\Delta E$ (kJ/mol)} \\ \hline
\textbf{Unbiased}  & 0.60  &  -   & -    & -    & -         & -       \\ \hline
\textbf{HLDA}      &   1.74  & 0.82 & 2.03 &2.31 & 18  &   10   \\ \hline
\textbf{Deep LDA}  &   1.89  & 0.97 & 1.89 &1.67 &  91  &   12   \\ \hline
\textbf{Deep TICA} &  1.28 &  0.96   & 1.38 &1.82 & 46 & 10     \\ \hline

\end{tabular}
\end{table*}

\begin{figure}[H]
    \centering
    \includegraphics[width=1\textwidth]{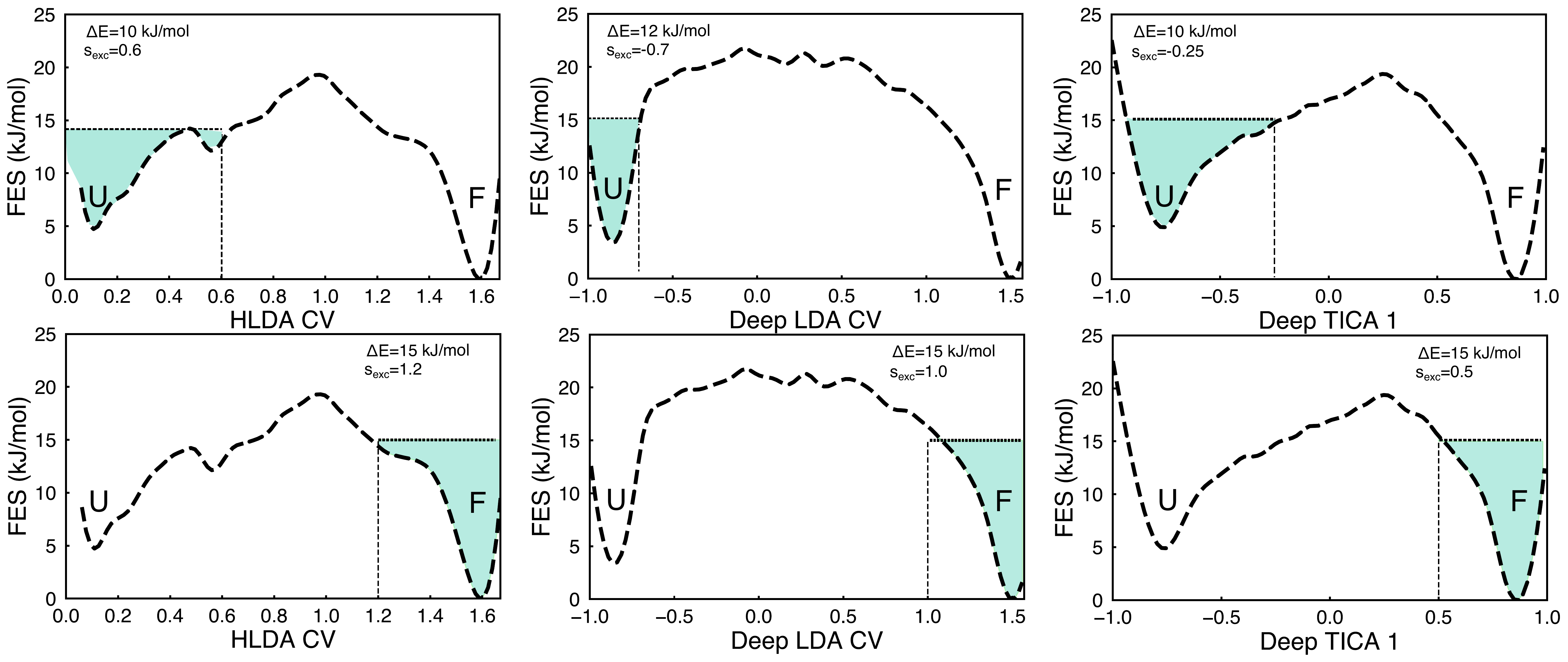} 
    \caption{Folding and Unfolding of chignolin mini-protein using OPES$_\mathrm{f}$ simulations. Panels show the free energy surface along three machine learning based collective variables: HLDA, Deep-LDA and Deep-TICA. The FES have been obtained from the long unbiased trajectory from Ref. \citenum{Lindorff-Larsen2011}. The region flooded with the OPES bias is shaded in light green for unfolded (upper) and folded (bottom) states.}
    \label{fig:chignolin_unfolding}
\end{figure}

The folding timescales have also been computed using the OPES$_\mathrm{f}$ procedure, and the results are in agreement with the unbiased estimate (within 2-3 factor) for all three choice of collective variables. The results as well as the simulation parameters are summarized in Table \ref{tab:Folding} and Fig. \ref{fig:chignolin_unfolding}  Similar to the unfolding simulations, we observe that the deep learning based collective variables perform better than the HLDA CV in terms of both efficiency and accuracy. Although the $\tau$ value for the HLDA and Deep-LDA are comparable, the p-value and acceleration factor of the deep-LDA calculations are clearly better. It should be noted that the optimum $\Delta E$ for the forward and the backward transitions can be different. For example, in chignolin, we used a lower $\Delta E$ for the folding process compared to that for the unfolding. This is necessary because the activation barrier for the folding process is relatively lower due to the thermodynamic stability of the folded structure.

\section{Discussions}
Based on the observations made in the current work, we recommend the following approach for calculating kinetics of a new system using OPES flooding approach.

\begin{itemize}
    \item If possible, one should calculate the free energy surface for the system or obtain it from the literature, if already published. This could be done only for a limited set of systems. 
    
    For a completely new system, calculating a converged free energy landscape can be difficult and expensive. In such situation, the FES can be computed only for the initial state basin and the TS region by applying a half harmonic wall to prevent exploration in the product region. (See Supporting information for an example on Chignolin). 
    
    \item From the FES (or partial FES), the approximate barrier height and the location of the transition state should be estimated. The $\Delta E$ for OPES$_\mathrm{f}$  has to be less than the barrier height. Being more conservative provides accurate kinetics, but acceleration will be small.
    
    The $\chi_{\textrm{exc}}(s)$ should be chosen appropriately to avoid biasing the transition state (See Fig. \ref{fig:chignolin_unfolding} for example). It is recommended to choose the $\Delta E$ to be approximately equal to or less than the value of the free energy surface at CV $s = s_\textrm{exc}$. Otherwise, the biasing potential can create an unphysical minimum to trap the system before transition, leading to an overestimation of the mean first passage time.
    
    \item If obtaining any kind of FES is very difficult due to the nature of the system and/or the CV, one might perform a few sets of OPES flooding simulations with decreasing $\Delta E$ starting from a very high value, and monitor the convergence of bias in the initial state basin before the transition takes place. The highest possible $\Delta E$ that consistently leads to converged OPES bias should be the best choice. \textcolor{red}{The additional computational cost for these test simulations will be minimal in comparison to the original OPES$_f$ simulations for two reasons. First, the simulation time is reduced by orders of magnitude when a higher value of $\Delta E$ is used. For example, we show in SI Table S9 that increasing $\Delta E$ by 5 kJ/mol results in an increase of the acceleration factor by approximately a factor of 5 for the alanine dipeptide system. Second, only a handful of test simulations, at each $\Delta E$, can provide an idea of convergence of the bias vs. time, before conducting multiple (usually 10 or more) flooding simulations at the optimized $\Delta E$ for computing converged rates.}
    
    \item Unlike infrequent metadynamics, it is not necessary to have a long time interval between the deposition of Gaussian kernel. In the alanine dipeptide system, for example, an interval of 500 time steps provided a reasonably accurate result. Using a very short interval can drive the system out of equilibrium, leading to errors in the estimated rates. A longer pace does not affect the quality of the results, but the computational cost increases unnecessarily (supporting information, Fig. S7). 
    
    \item The width of the Gaussian kernel estimated automatically by the OPES module can be too high for the purpose of rate calculation. For OPES flooding, one needs slightly narrower kernels so that the tails of the Gaussian kernels do not extend into the transition state. For our alanine dipeptide system, a kernel width that is 2-3 times smaller than the one obtained from the initial OPES run provided optimum results. (supporting information, Fig. S8).
    
    It should be noted that this parameter only indicates the width of the first Gaussian kernel deposited. The OPES method uses a data driven technique to optimize the width of the kernels as the simulation proceeds. For more details see Ref. \citenum{Invernizzi2020}
    
    \item The transition trajectories should be stopped once they reach the final state. One should make sure that the system spends a few hundred to a few thousand time steps (depending on the transition timescales) in the final state before stopping the trajectory, to ensure that the system has reached a new metastable state and the observed transition is not just an artifact of the CV. When using a suboptimal CV, one should consider using more than one CVs to precisely define the final state to make sure that the trajectories are stopped if and only if they reach the correct final state. For example, we used both $\phi$ and $\psi$ torsion angles to define the final state for alanine dipeptide when we biased the $\psi$ CV which is unable to properly distinguish the two state. 
    
    \textcolor{red}{Nevertheless, we want to emphasize that the above discussion is applicable to suboptimal CVs which can approximately distinguish the initial and the final state. When the CV is so poor that it cannot properly distinguish the two metastable states, for example the y coordinate for the $\theta$ = 54$^{\circ}$ case of the Wolfe-Quapp potential (Fig. \ref{fig:2d_model_results}b), it will no longer be possible to define the boundary of the final state for truncating the trajectories. Any enhanced sampling method with this type of CV is futile and no kinetic or thermodynamic information can be recovered unless the CV is improved. }
    
    \item Even after following all these procedures, if one finds that the transitions are too rare (low acceleration factor), one should consider improving the CV. It is possible to use a higher $\Delta E$ when using a better CV, because the effective barrier height will be higher if the CV can distinguish the TS correctly. In this vein, the newly designed deep learning based collective variables including Deep-LDA \cite{Bonati2020} and Deep-TICA \cite{Bonati2021} can be of significant help. 
    
    \item If the deposited bias exceeds the barrier height and (or) the transition state is corrupted by the OPES kernels, the quality of kinetics cannot be assessed reliably by the KS test alone. It is important to carefully check the convergence of bias in the initial state basin before such analysis is performed. 
\end{itemize}

\section{Conclusions}
We propose a variant of the On-the-fly Probability Enhanced Sampling (OPES) for calculating the kinetics of rare event processes from atomistic MD simulation. For model systems and explicit solvent protein molecules, we carefully tested the utility and limits of our technique, which we refer to as OPES flooding. 
Our results demonstrate that the most important consideration in any enhanced sampling approach for kinetics, including infrequent metadynamics and OPES flooding, is that the deposited bias must become quasistatic in the initial stated basin so that the ensemble average in the expression of acceleration factor can be correctly evaluated. Infrequent metadynamics approximates this condition by employing very low bias deposition rate, which also decreases the likelihood of biasing the transition state. 
The OPES$_\mathrm{f}$ approach improves upon infrequent metadynamics by allowing the user to control the amount and location of the bias deposition through the $\Delta E$ and $\chi_{\textrm{exc}}(s)$ parameters. Unlike the infrequent metadynamics a large interval between the deposition of successive Gaussian kernels is not necessary, leading to an increase in the speed-up of barrier crossing transitions. In metadynamics the bias increases constantly, making it difficult to directly control the amount of bias deposition.
Whereas, in OPES flooding, as long as $\Delta E$ does not exceed the true activation energy barrier and the $\chi_{\textrm{exc}}(s)$ can ensure that the transition state is bias free, accurate kinetics can be predicted. Moreover, the computed timescales are reasonably accurate even when the CV is suboptimal, although improving the quality of the CV will increase the efficiency of the simulation. The underlying reason is that the FES along a better CV will potentially have a higher effective free energy barrier allowing for the use of a higher $\Delta E$ and consequently increasing the acceleration factor. Thus, with the correct choice of a handful of parameters, the user has direct control over the accuracy of the kinetic results obtained from OPES$_\mathrm{f}$ simulations. Besides the results presented here mostly for deductive purpose, the interested reader can find a more real life application of OPES$_\mathrm{f}$ for calculating millisecond timescale dissociation kinetics of a protein-ligand complex in Ref.~\citenum{Ansari2022Water}. Considering the additional advantages over traditional enhanced sampling methods, and the growing need for better methods for predicting kinetics in molecular systems, we expect that OPES flooding technique will find a wide range of applications in molecular biophysics, chemistry, and material science.

\section*{Data Availability}
The OPES flooding approach is implemented in PLUMED from version 2.8, but for the simulations the latest PLUMED 2.9 was used. The input files and the analysis scripts for all simulations performed in this work are available in github (\url{https://github.com/dhimanray/OPES-Flooding}). The input files are also available through the PLUMED NEST repository\cite{nest}. 

\section*{Supplementary Information Available}
See the supplementary information for discussions on the dimensionality of the excluded region, the generation of an approximate FES for estimating parameters for OPES flooding simulation, details of construction of deep learning CVs, data tables (Tab. S1-S12) for kinetics and KS test results, and supplementary figures (Fig. S1-S9)

\section*{Acknowledgements}
M.I. acknowledges support from the Swiss National Science Foundation through an Early Postdoc.Mobility fellowship. The authors thank Davide Mandelli, Umberto Raucci, Ana Borrego-S{\'a}nchez, Jayashrita Debnath, Luigi Bonati, and Andrea Rizzi for helpful discussions. The authors thank D.E. Shaw Research for sharing the trajectories of Chignolin protein from Ref. \citenum{Lindorff-Larsen2011}. The authors declare no competing financial interest.


\bibliography{references,ref1}{}

\providecommand{\latin}[1]{#1}
\makeatletter
\providecommand{\doi}
  {\begingroup\let\do\@makeother\dospecials
  \catcode`\{=1 \catcode`\}=2 \doi@aux}
\providecommand{\doi@aux}[1]{\endgroup\texttt{#1}}
\makeatother
\providecommand*\mcitethebibliography{\thebibliography}
\csname @ifundefined\endcsname{endmcitethebibliography}
  {\let\endmcitethebibliography\endthebibliography}{}
\begin{mcitethebibliography}{49}
\providecommand*\natexlab[1]{#1}
\providecommand*\mciteSetBstSublistMode[1]{}
\providecommand*\mciteSetBstMaxWidthForm[2]{}
\providecommand*\mciteBstWouldAddEndPuncttrue
  {\def\EndOfBibitem{\unskip.}}
\providecommand*\mciteBstWouldAddEndPunctfalse
  {\let\EndOfBibitem\relax}
\providecommand*\mciteSetBstMidEndSepPunct[3]{}
\providecommand*\mciteSetBstSublistLabelBeginEnd[3]{}
\providecommand*\EndOfBibitem{}
\mciteSetBstSublistMode{f}
\mciteSetBstMaxWidthForm{subitem}{(\alph{mcitesubitemcount})}
\mciteSetBstSublistLabelBeginEnd
  {\mcitemaxwidthsubitemform\space}
  {\relax}
  {\relax}

\bibitem[Ahmad \latin{et~al.}(2022)Ahmad, Rizzi, Capelli, Mandelli, Lyu, and
  Carloni]{Ahmad2022}
Ahmad,~K.; Rizzi,~A.; Capelli,~R.; Mandelli,~D.; Lyu,~W.; Carloni,~P.
  {Enhanced-Sampling Simulations for the Estimation of Ligand Binding Kinetics:
  Current Status and Perspective}. \emph{Frontiers in Molecular Biosciences}
  \textbf{2022}, \emph{9}, 1--17\relax
\mciteBstWouldAddEndPuncttrue
\mciteSetBstMidEndSepPunct{\mcitedefaultmidpunct}
{\mcitedefaultendpunct}{\mcitedefaultseppunct}\relax
\EndOfBibitem
\bibitem[Palacio-Rodriguez \latin{et~al.}(2021)Palacio-Rodriguez, Vroylandt,
  Stelzl, Pietrucci, Hummer, and Cossio]{Palacio-Rodriguez2021}
Palacio-Rodriguez,~K.; Vroylandt,~H.; Stelzl,~L.~S.; Pietrucci,~F.; Hummer,~G.;
  Cossio,~P. {Transition rates, survival probabilities, and quality of bias
  from time-dependent biased simulations}. \textbf{2021}, 1--11\relax
\mciteBstWouldAddEndPuncttrue
\mciteSetBstMidEndSepPunct{\mcitedefaultmidpunct}
{\mcitedefaultendpunct}{\mcitedefaultseppunct}\relax
\EndOfBibitem
\bibitem[Donati and Keller(2018)Donati, and Keller]{donati2018girsanov}
Donati,~L.; Keller,~B.~G. Girsanov reweighting for metadynamics simulations.
  \emph{The Journal of chemical physics} \textbf{2018}, \emph{149},
  072335\relax
\mciteBstWouldAddEndPuncttrue
\mciteSetBstMidEndSepPunct{\mcitedefaultmidpunct}
{\mcitedefaultendpunct}{\mcitedefaultseppunct}\relax
\EndOfBibitem
\bibitem[Bolhuis \latin{et~al.}(2002)Bolhuis, Chandler, Dellago, and
  Geissler]{Bolhuis2002}
Bolhuis,~P.~G.; Chandler,~D.; Dellago,~C.; Geissler,~P.~L. {Transition Path
  Sampling: Throwing Ropes Over Rough Mountain Passes, in the Dark}.
  \emph{Annual Review of Physical Chemistry} \textbf{2002}, \emph{53},
  291--318\relax
\mciteBstWouldAddEndPuncttrue
\mciteSetBstMidEndSepPunct{\mcitedefaultmidpunct}
{\mcitedefaultendpunct}{\mcitedefaultseppunct}\relax
\EndOfBibitem
\bibitem[Huber and Kim(1996)Huber, and Kim]{huber1996weighted}
Huber,~G.~A.; Kim,~S. Weighted-ensemble Brownian dynamics simulations for
  protein association reactions. \emph{Biophysical journal} \textbf{1996},
  \emph{70}, 97--110\relax
\mciteBstWouldAddEndPuncttrue
\mciteSetBstMidEndSepPunct{\mcitedefaultmidpunct}
{\mcitedefaultendpunct}{\mcitedefaultseppunct}\relax
\EndOfBibitem
\bibitem[Zhang \latin{et~al.}(2010)Zhang, Jasnow, and
  Zuckerman]{zhang2010weighted}
Zhang,~B.~W.; Jasnow,~D.; Zuckerman,~D.~M. The “weighted ensemble” path
  sampling method is statistically exact for a broad class of stochastic
  processes and binning procedures. \emph{The Journal of chemical physics}
  \textbf{2010}, \emph{132}, 054107\relax
\mciteBstWouldAddEndPuncttrue
\mciteSetBstMidEndSepPunct{\mcitedefaultmidpunct}
{\mcitedefaultendpunct}{\mcitedefaultseppunct}\relax
\EndOfBibitem
\bibitem[Zuckerman and Chong(2017)Zuckerman, and Chong]{zuckerman2017weighted}
Zuckerman,~D.~M.; Chong,~L.~T. Weighted ensemble simulation: review of
  methodology, applications, and software. \emph{Annual review of biophysics}
  \textbf{2017}, \emph{46}, 43\relax
\mciteBstWouldAddEndPuncttrue
\mciteSetBstMidEndSepPunct{\mcitedefaultmidpunct}
{\mcitedefaultendpunct}{\mcitedefaultseppunct}\relax
\EndOfBibitem
\bibitem[Faradjian and Elber(2004)Faradjian, and Elber]{Faradjian2004}
Faradjian,~A.~K.; Elber,~R. {Computing time scales from reaction coordinates by
  milestoning}. \emph{The Journal of Chemical Physics} \textbf{2004},
  \emph{120}, 10880--10889\relax
\mciteBstWouldAddEndPuncttrue
\mciteSetBstMidEndSepPunct{\mcitedefaultmidpunct}
{\mcitedefaultendpunct}{\mcitedefaultseppunct}\relax
\EndOfBibitem
\bibitem[West \latin{et~al.}(2007)West, Elber, and Shalloway]{West2007}
West,~A. M.~A.; Elber,~R.; Shalloway,~D. {Extending molecular dynamics time
  scales with milestoning: Example of complex kinetics in a solvated peptide}.
  \emph{The Journal of Chemical Physics} \textbf{2007}, \emph{126},
  145104\relax
\mciteBstWouldAddEndPuncttrue
\mciteSetBstMidEndSepPunct{\mcitedefaultmidpunct}
{\mcitedefaultendpunct}{\mcitedefaultseppunct}\relax
\EndOfBibitem
\bibitem[Allen \latin{et~al.}(2006)Allen, Frenkel, and ten
  Wolde]{allen2006simulating}
Allen,~R.~J.; Frenkel,~D.; ten Wolde,~P.~R. Simulating rare events in
  equilibrium or nonequilibrium stochastic systems. \emph{The Journal of
  chemical physics} \textbf{2006}, \emph{124}, 024102\relax
\mciteBstWouldAddEndPuncttrue
\mciteSetBstMidEndSepPunct{\mcitedefaultmidpunct}
{\mcitedefaultendpunct}{\mcitedefaultseppunct}\relax
\EndOfBibitem
\bibitem[Ray and Andricioaei(2020)Ray, and
  Andricioaei]{Ray2020WeightedSimulations}
Ray,~D.; Andricioaei,~I. {Weighted ensemble milestoning (WEM): A combined
  approach for rare event simulations}. \emph{The Journal of Chemical Physics}
  \textbf{2020}, \emph{152}, 234114\relax
\mciteBstWouldAddEndPuncttrue
\mciteSetBstMidEndSepPunct{\mcitedefaultmidpunct}
{\mcitedefaultendpunct}{\mcitedefaultseppunct}\relax
\EndOfBibitem
\bibitem[Ray \latin{et~al.}(2022)Ray, Stone, and Andricioaei]{Ray2022}
Ray,~D.; Stone,~S.~E.; Andricioaei,~I. {Markovian Weighted Ensemble Milestoning
  (M-WEM): Long-Time Kinetics from Short Trajectories}. \emph{Journal of
  Chemical Theory and Computation} \textbf{2022}, \emph{18}, 79--95\relax
\mciteBstWouldAddEndPuncttrue
\mciteSetBstMidEndSepPunct{\mcitedefaultmidpunct}
{\mcitedefaultendpunct}{\mcitedefaultseppunct}\relax
\EndOfBibitem
\bibitem[Grubm{\"{u}}ller(1995)]{Grubmuller1995}
Grubm{\"{u}}ller,~H. {Predicting slow structural transitions in macromolecular
  systems: Conformational flooding}. \emph{Physical Review E} \textbf{1995},
  \emph{52}, 2893--2906\relax
\mciteBstWouldAddEndPuncttrue
\mciteSetBstMidEndSepPunct{\mcitedefaultmidpunct}
{\mcitedefaultendpunct}{\mcitedefaultseppunct}\relax
\EndOfBibitem
\bibitem[Voter(1997)]{Voter1997}
Voter,~A.~F. {A method for accelerating the molecular dynamics simulation of
  infrequent events}. \emph{The Journal of Chemical Physics} \textbf{1997},
  \emph{106}, 4665--4677\relax
\mciteBstWouldAddEndPuncttrue
\mciteSetBstMidEndSepPunct{\mcitedefaultmidpunct}
{\mcitedefaultendpunct}{\mcitedefaultseppunct}\relax
\EndOfBibitem
\bibitem[Laio and Parrinello(2002)Laio, and Parrinello]{Laio2002}
Laio,~A.; Parrinello,~M. {Escaping free-energy minima}. \emph{Proceedings of
  the National Academy of Sciences} \textbf{2002}, \emph{99},
  12562--12566\relax
\mciteBstWouldAddEndPuncttrue
\mciteSetBstMidEndSepPunct{\mcitedefaultmidpunct}
{\mcitedefaultendpunct}{\mcitedefaultseppunct}\relax
\EndOfBibitem
\bibitem[Barducci \latin{et~al.}(2008)Barducci, Bussi, and
  Parrinello]{Barducci2008}
Barducci,~A.; Bussi,~G.; Parrinello,~M. {Well-Tempered Metadynamics: A Smoothly
  Converging and Tunable Free-Energy Method}. \emph{Physical Review Letters}
  \textbf{2008}, \emph{100}, 020603\relax
\mciteBstWouldAddEndPuncttrue
\mciteSetBstMidEndSepPunct{\mcitedefaultmidpunct}
{\mcitedefaultendpunct}{\mcitedefaultseppunct}\relax
\EndOfBibitem
\bibitem[Tiwary and Parrinello(2013)Tiwary, and Parrinello]{Tiwary2013}
Tiwary,~P.; Parrinello,~M. {From Metadynamics to Dynamics}. \emph{Physical
  Review Letters} \textbf{2013}, \emph{111}, 230602\relax
\mciteBstWouldAddEndPuncttrue
\mciteSetBstMidEndSepPunct{\mcitedefaultmidpunct}
{\mcitedefaultendpunct}{\mcitedefaultseppunct}\relax
\EndOfBibitem
\bibitem[Casasnovas \latin{et~al.}(2017)Casasnovas, Limongelli, Tiwary,
  Carloni, and Parrinello]{Casasnovas2017}
Casasnovas,~R.; Limongelli,~V.; Tiwary,~P.; Carloni,~P.; Parrinello,~M.
  {Unbinding Kinetics of a p38 MAP Kinase Type II Inhibitor from Metadynamics
  Simulations}. \emph{Journal of the American Chemical Society} \textbf{2017},
  \emph{139}, 4780--4788\relax
\mciteBstWouldAddEndPuncttrue
\mciteSetBstMidEndSepPunct{\mcitedefaultmidpunct}
{\mcitedefaultendpunct}{\mcitedefaultseppunct}\relax
\EndOfBibitem
\bibitem[Tiwary \latin{et~al.}(2015)Tiwary, Limongelli, Salvalaglio, and
  Parrinello]{Tiwary2015Trypsin}
Tiwary,~P.; Limongelli,~V.; Salvalaglio,~M.; Parrinello,~M. Kinetics of
  protein-ligand unbinding: Predicting pathways, rates, and rate-limiting
  steps. \emph{Proceedings of the National Academy of Sciences} \textbf{2015},
  \emph{112}, E386--E391\relax
\mciteBstWouldAddEndPuncttrue
\mciteSetBstMidEndSepPunct{\mcitedefaultmidpunct}
{\mcitedefaultendpunct}{\mcitedefaultseppunct}\relax
\EndOfBibitem
\bibitem[Ribeiro \latin{et~al.}(2018)Ribeiro, Bravo, Wang, and
  Tiwary]{ribeiro2018reweighted}
Ribeiro,~J. M.~L.; Bravo,~P.; Wang,~Y.; Tiwary,~P. Reweighted autoencoded
  variational Bayes for enhanced sampling (RAVE). \emph{The Journal of chemical
  physics} \textbf{2018}, \emph{149}, 072301\relax
\mciteBstWouldAddEndPuncttrue
\mciteSetBstMidEndSepPunct{\mcitedefaultmidpunct}
{\mcitedefaultendpunct}{\mcitedefaultseppunct}\relax
\EndOfBibitem
\bibitem[Rizzi \latin{et~al.}(2019)Rizzi, Polino, Sicilia, Russo, and
  Parrinello]{rizzi2019onset}
Rizzi,~V.; Polino,~D.; Sicilia,~E.; Russo,~N.; Parrinello,~M. The onset of
  dehydrogenation in solid ammonia borane: An ab initio metadynamics study.
  \emph{Angewandte Chemie} \textbf{2019}, \emph{131}, 4016--4020\relax
\mciteBstWouldAddEndPuncttrue
\mciteSetBstMidEndSepPunct{\mcitedefaultmidpunct}
{\mcitedefaultendpunct}{\mcitedefaultseppunct}\relax
\EndOfBibitem
\bibitem[Mondal \latin{et~al.}(2018)Mondal, Ahalawat, Pandit, Kay, and
  Vallurupalli]{mondal2018atomic}
Mondal,~J.; Ahalawat,~N.; Pandit,~S.; Kay,~L.~E.; Vallurupalli,~P. Atomic
  resolution mechanism of ligand binding to a solvent inaccessible cavity in T4
  lysozyme. \emph{PLoS computational biology} \textbf{2018}, \emph{14},
  e1006180\relax
\mciteBstWouldAddEndPuncttrue
\mciteSetBstMidEndSepPunct{\mcitedefaultmidpunct}
{\mcitedefaultendpunct}{\mcitedefaultseppunct}\relax
\EndOfBibitem
\bibitem[Kulkarni and S{\"o}derhjelm(2021)Kulkarni, and
  S{\"o}derhjelm]{Kulkarni2021BPTI}
Kulkarni,~M.; S{\"o}derhjelm,~P. Free Energy Landscape and Rate Estimation of
  the Aromatic Ring Flips in Basic Pancreatic Trypsin Inhibitor Using
  Metadynamics. \emph{bioRxiv} \textbf{2021}, \relax
\mciteBstWouldAddEndPunctfalse
\mciteSetBstMidEndSepPunct{\mcitedefaultmidpunct}
{}{\mcitedefaultseppunct}\relax
\EndOfBibitem
\bibitem[Yang \latin{et~al.}(2022)Yang, Bonati, Polino, and
  Parrinello]{yang2022using}
Yang,~M.; Bonati,~L.; Polino,~D.; Parrinello,~M. Using metadynamics to build
  neural network potentials for reactive events: the case of urea decomposition
  in water. \emph{Catalysis Today} \textbf{2022}, \emph{387}, 143--149\relax
\mciteBstWouldAddEndPuncttrue
\mciteSetBstMidEndSepPunct{\mcitedefaultmidpunct}
{\mcitedefaultendpunct}{\mcitedefaultseppunct}\relax
\EndOfBibitem
\bibitem[Dickson(2019)]{Dickson2019}
Dickson,~B.~M. {Erroneous Rates and False Statistical Confirmations from
  Infrequent Metadynamics and Other Equivalent Violations of the Hyperdynamics
  Paradigm}. \emph{Journal of Chemical Theory and Computation} \textbf{2019},
  \emph{15}, 78--83\relax
\mciteBstWouldAddEndPuncttrue
\mciteSetBstMidEndSepPunct{\mcitedefaultmidpunct}
{\mcitedefaultendpunct}{\mcitedefaultseppunct}\relax
\EndOfBibitem
\bibitem[McCarty \latin{et~al.}(2015)McCarty, Valsson, Tiwary, and
  Parrinello]{McCarty2015}
McCarty,~J.; Valsson,~O.; Tiwary,~P.; Parrinello,~M. {Variationally Optimized
  Free-Energy Flooding for Rate Calculation}. \emph{Physical Review Letters}
  \textbf{2015}, \emph{115}, 070601\relax
\mciteBstWouldAddEndPuncttrue
\mciteSetBstMidEndSepPunct{\mcitedefaultmidpunct}
{\mcitedefaultendpunct}{\mcitedefaultseppunct}\relax
\EndOfBibitem
\bibitem[Debnath and Parrinello(2020)Debnath, and Parrinello]{Debnath2020}
Debnath,~J.; Parrinello,~M. {Gaussian Mixture-Based Enhanced Sampling for
  Statics and Dynamics}. \emph{The Journal of Physical Chemistry Letters}
  \textbf{2020}, \emph{11}, 5076--5080\relax
\mciteBstWouldAddEndPuncttrue
\mciteSetBstMidEndSepPunct{\mcitedefaultmidpunct}
{\mcitedefaultendpunct}{\mcitedefaultseppunct}\relax
\EndOfBibitem
\bibitem[Piccini \latin{et~al.}(2017)Piccini, McCarty, Valsson, and
  Parrinello]{Piccini2017}
Piccini,~G.; McCarty,~J.~J.; Valsson,~O.; Parrinello,~M. {Variational Flooding
  Study of a S N 2 Reaction}. \emph{The Journal of Physical Chemistry Letters}
  \textbf{2017}, \emph{8}, 580--583\relax
\mciteBstWouldAddEndPuncttrue
\mciteSetBstMidEndSepPunct{\mcitedefaultmidpunct}
{\mcitedefaultendpunct}{\mcitedefaultseppunct}\relax
\EndOfBibitem
\bibitem[Palazzesi \latin{et~al.}(2017)Palazzesi, Valsson, and
  Parrinello]{palazzesi2017conformational}
Palazzesi,~F.; Valsson,~O.; Parrinello,~M. Conformational entropy as collective
  variable for proteins. \emph{The journal of physical chemistry letters}
  \textbf{2017}, \emph{8}, 4752--4756\relax
\mciteBstWouldAddEndPuncttrue
\mciteSetBstMidEndSepPunct{\mcitedefaultmidpunct}
{\mcitedefaultendpunct}{\mcitedefaultseppunct}\relax
\EndOfBibitem
\bibitem[Debnath and Parrinello(2022)Debnath, and
  Parrinello]{debnath2022computing}
Debnath,~J.; Parrinello,~M. Computing Rates and Understanding Unbinding
  Mechanisms in Host--Guest Systems. \emph{Journal of Chemical Theory and
  Computation} \textbf{2022}, \emph{18}, 1314--1319\relax
\mciteBstWouldAddEndPuncttrue
\mciteSetBstMidEndSepPunct{\mcitedefaultmidpunct}
{\mcitedefaultendpunct}{\mcitedefaultseppunct}\relax
\EndOfBibitem
\bibitem[Invernizzi and Parrinello(2020)Invernizzi, and
  Parrinello]{Invernizzi2020}
Invernizzi,~M.; Parrinello,~M. {Rethinking Metadynamics: From Bias Potentials
  to Probability Distributions}. \emph{The Journal of Physical Chemistry
  Letters} \textbf{2020}, \emph{11}, 2731--2736\relax
\mciteBstWouldAddEndPuncttrue
\mciteSetBstMidEndSepPunct{\mcitedefaultmidpunct}
{\mcitedefaultendpunct}{\mcitedefaultseppunct}\relax
\EndOfBibitem
\bibitem[Invernizzi \latin{et~al.}(2020)Invernizzi, Piaggi, and
  Parrinello]{Invernizzi2020d}
Invernizzi,~M.; Piaggi,~P.~M.; Parrinello,~M. {Unified Approach to Enhanced
  Sampling}. \emph{Physical Review X} \textbf{2020}, \emph{10}, 041034\relax
\mciteBstWouldAddEndPuncttrue
\mciteSetBstMidEndSepPunct{\mcitedefaultmidpunct}
{\mcitedefaultendpunct}{\mcitedefaultseppunct}\relax
\EndOfBibitem
\bibitem[Invernizzi and Parrinello(2022)Invernizzi, and
  Parrinello]{Invernizzi2022}
Invernizzi,~M.; Parrinello,~M. {Exploration vs Convergence Speed in
  Adaptive-Bias Enhanced Sampling}. \emph{Journal of Chemical Theory and
  Computation} \textbf{2022}, \emph{18}, 3988--3996\relax
\mciteBstWouldAddEndPuncttrue
\mciteSetBstMidEndSepPunct{\mcitedefaultmidpunct}
{\mcitedefaultendpunct}{\mcitedefaultseppunct}\relax
\EndOfBibitem
\bibitem[Hamelberg \latin{et~al.}(2004)Hamelberg, Mongan, and
  McCammon]{Hamelberg2004}
Hamelberg,~D.; Mongan,~J.; McCammon,~J.~A. {Accelerated molecular dynamics: A
  promising and efficient simulation method for biomolecules}. \emph{The
  Journal of Chemical Physics} \textbf{2004}, \emph{120}, 11919--11929\relax
\mciteBstWouldAddEndPuncttrue
\mciteSetBstMidEndSepPunct{\mcitedefaultmidpunct}
{\mcitedefaultendpunct}{\mcitedefaultseppunct}\relax
\EndOfBibitem
\bibitem[Salvalaglio \latin{et~al.}(2014)Salvalaglio, Tiwary, and
  Parrinello]{Salvalaglio2014}
Salvalaglio,~M.; Tiwary,~P.; Parrinello,~M. {Assessing the Reliability of the
  Dynamics Reconstructed from Metadynamics}. \emph{Journal of Chemical Theory
  and Computation} \textbf{2014}, \emph{10}, 1420--1425\relax
\mciteBstWouldAddEndPuncttrue
\mciteSetBstMidEndSepPunct{\mcitedefaultmidpunct}
{\mcitedefaultendpunct}{\mcitedefaultseppunct}\relax
\EndOfBibitem
\bibitem[Invernizzi and Parrinello(2019)Invernizzi, and
  Parrinello]{Invernizzi2019}
Invernizzi,~M.; Parrinello,~M. {Making the Best of a Bad Situation: A
  Multiscale Approach to Free Energy Calculation}. \emph{Journal of Chemical
  Theory and Computation} \textbf{2019}, \relax
\mciteBstWouldAddEndPunctfalse
\mciteSetBstMidEndSepPunct{\mcitedefaultmidpunct}
{}{\mcitedefaultseppunct}\relax
\EndOfBibitem
\bibitem[Abraham \latin{et~al.}(2015)Abraham, Murtola, Schulz, P{\'{a}}ll,
  Smith, Hess, and Lindahl]{Abraham2015}
Abraham,~M.~J.; Murtola,~T.; Schulz,~R.; P{\'{a}}ll,~S.; Smith,~J.~C.;
  Hess,~B.; Lindahl,~E. {GROMACS: High performance molecular simulations
  through multi-level parallelism from laptops to supercomputers}.
  \emph{SoftwareX} \textbf{2015}, \emph{1-2}, 19--25\relax
\mciteBstWouldAddEndPuncttrue
\mciteSetBstMidEndSepPunct{\mcitedefaultmidpunct}
{\mcitedefaultendpunct}{\mcitedefaultseppunct}\relax
\EndOfBibitem
\bibitem[Tribello \latin{et~al.}(2014)Tribello, Bonomi, Branduardi, Camilloni,
  and Bussi]{Tribello2014}
Tribello,~G.~A.; Bonomi,~M.; Branduardi,~D.; Camilloni,~C.; Bussi,~G. {PLUMED
  2: New feathers for an old bird}. \emph{Computer Physics Communications}
  \textbf{2014}, \emph{185}, 604--613\relax
\mciteBstWouldAddEndPuncttrue
\mciteSetBstMidEndSepPunct{\mcitedefaultmidpunct}
{\mcitedefaultendpunct}{\mcitedefaultseppunct}\relax
\EndOfBibitem
\bibitem[Valsson \latin{et~al.}(2016)Valsson, Tiwary, and
  Parrinello]{Valsson2016}
Valsson,~O.; Tiwary,~P.; Parrinello,~M. {Enhancing Important Fluctuations: Rare
  Events and Metadynamics from a Conceptual Viewpoint}. \emph{Annual Review of
  Physical Chemistry} \textbf{2016}, \emph{67}, 159--184\relax
\mciteBstWouldAddEndPuncttrue
\mciteSetBstMidEndSepPunct{\mcitedefaultmidpunct}
{\mcitedefaultendpunct}{\mcitedefaultseppunct}\relax
\EndOfBibitem
\bibitem[Honda \latin{et~al.}(2008)Honda, Akiba, Kato, Sawada, Sekijima,
  Ishimura, Ooishi, Watanabe, Odahara, and Harata]{Honda2008}
Honda,~S.; Akiba,~T.; Kato,~Y.~S.; Sawada,~Y.; Sekijima,~M.; Ishimura,~M.;
  Ooishi,~A.; Watanabe,~H.; Odahara,~T.; Harata,~K. Crystal Structure of a
  Ten-Amino Acid Protein. \emph{Journal of the American Chemical Society}
  \textbf{2008}, \emph{130}, 15327--15331, PMID: 18950166\relax
\mciteBstWouldAddEndPuncttrue
\mciteSetBstMidEndSepPunct{\mcitedefaultmidpunct}
{\mcitedefaultendpunct}{\mcitedefaultseppunct}\relax
\EndOfBibitem
\bibitem[Lindorff-Larsen \latin{et~al.}(2011)Lindorff-Larsen, Piana, Dror, and
  Shaw]{Lindorff-Larsen2011}
Lindorff-Larsen,~K.; Piana,~S.; Dror,~R.~O.; Shaw,~D.~E. {How Fast-Folding
  Proteins Fold}. \emph{Science} \textbf{2011}, \emph{334}, 517--520\relax
\mciteBstWouldAddEndPuncttrue
\mciteSetBstMidEndSepPunct{\mcitedefaultmidpunct}
{\mcitedefaultendpunct}{\mcitedefaultseppunct}\relax
\EndOfBibitem
\bibitem[Piana \latin{et~al.}(2011)Piana, Lindorff-Larsen, and
  Shaw]{piana2011robust}
Piana,~S.; Lindorff-Larsen,~K.; Shaw,~D.~E. How robust are protein folding
  simulations with respect to force field parameterization? \emph{Biophysical
  journal} \textbf{2011}, \emph{100}, L47--L49\relax
\mciteBstWouldAddEndPuncttrue
\mciteSetBstMidEndSepPunct{\mcitedefaultmidpunct}
{\mcitedefaultendpunct}{\mcitedefaultseppunct}\relax
\EndOfBibitem
\bibitem[MacKerell~Jr \latin{et~al.}(1998)MacKerell~Jr, Bashford, Bellott,
  Dunbrack~Jr, Evanseck, Field, Fischer, Gao, Guo, Ha, \latin{et~al.}
  others]{mackerell1998all}
MacKerell~Jr,~A.~D.; Bashford,~D.; Bellott,~M.; Dunbrack~Jr,~R.~L.;
  Evanseck,~J.~D.; Field,~M.~J.; Fischer,~S.; Gao,~J.; Guo,~H.; Ha,~S.,
  \latin{et~al.}  All-atom empirical potential for molecular modeling and
  dynamics studies of proteins. \emph{The journal of physical chemistry B}
  \textbf{1998}, \emph{102}, 3586--3616\relax
\mciteBstWouldAddEndPuncttrue
\mciteSetBstMidEndSepPunct{\mcitedefaultmidpunct}
{\mcitedefaultendpunct}{\mcitedefaultseppunct}\relax
\EndOfBibitem
\bibitem[Mendels \latin{et~al.}(2018)Mendels, Piccini, Brotzakis, Yang, and
  Parrinello]{Mendels2018a}
Mendels,~D.; Piccini,~G.; Brotzakis,~Z.~F.; Yang,~Y.~I.; Parrinello,~M.
  {Folding a small protein using harmonic linear discriminant analysis}.
  \emph{The Journal of Chemical Physics} \textbf{2018}, \emph{149},
  194113\relax
\mciteBstWouldAddEndPuncttrue
\mciteSetBstMidEndSepPunct{\mcitedefaultmidpunct}
{\mcitedefaultendpunct}{\mcitedefaultseppunct}\relax
\EndOfBibitem
\bibitem[Bonati \latin{et~al.}(2020)Bonati, Rizzi, and Parrinello]{Bonati2020}
Bonati,~L.; Rizzi,~V.; Parrinello,~M. {Data-Driven Collective Variables for
  Enhanced Sampling}. \emph{The Journal of Physical Chemistry Letters}
  \textbf{2020}, \emph{11}, 2998--3004\relax
\mciteBstWouldAddEndPuncttrue
\mciteSetBstMidEndSepPunct{\mcitedefaultmidpunct}
{\mcitedefaultendpunct}{\mcitedefaultseppunct}\relax
\EndOfBibitem
\bibitem[Bonati \latin{et~al.}(2021)Bonati, Piccini, and
  Parrinello]{Bonati2021}
Bonati,~L.; Piccini,~G.; Parrinello,~M. {Deep learning the slow modes for rare
  events sampling}. \emph{Proceedings of the National Academy of Sciences}
  \textbf{2021}, \emph{118}\relax
\mciteBstWouldAddEndPuncttrue
\mciteSetBstMidEndSepPunct{\mcitedefaultmidpunct}
{\mcitedefaultendpunct}{\mcitedefaultseppunct}\relax
\EndOfBibitem
\bibitem[Ansari \latin{et~al.}(2022)Ansari, Rizzi, and
  Parrinello]{Ansari2022Water}
Ansari,~N.; Rizzi,~V.; Parrinello,~M. Water regulates the residence time of
  Benzamidine in Trypsin. 2022; \url{https://arxiv.org/abs/2204.05572}\relax
\mciteBstWouldAddEndPuncttrue
\mciteSetBstMidEndSepPunct{\mcitedefaultmidpunct}
{\mcitedefaultendpunct}{\mcitedefaultseppunct}\relax
\EndOfBibitem
\bibitem[{The PLUMED consortium}(2019)]{nest}
{The PLUMED consortium}, {Promoting transparency and reproducibility in
  enhanced molecular simulations}. \emph{Nature Methods} \textbf{2019},
  \emph{16}, 670--673\relax
\mciteBstWouldAddEndPuncttrue
\mciteSetBstMidEndSepPunct{\mcitedefaultmidpunct}
{\mcitedefaultendpunct}{\mcitedefaultseppunct}\relax
\EndOfBibitem
\end{mcitethebibliography}


\end{document}